\documentclass[conference]{IEEEtran}
\IEEEoverridecommandlockouts

\usepackage{cite}
\usepackage{graphicx}
\usepackage[table,xcdraw]{xcolor}
\usepackage{url}
\usepackage{amsmath,amssymb,amsfonts}
\usepackage{algorithmic}
\usepackage{hyperref}
\usepackage{multirow}
\usepackage{textcomp}

\def\BibTeX{{\rm B\kern-.05em{\sc i\kern-.025em b}\kern-.08em
    T\kern-.1667em\lower.7ex\hbox{E}\kern-.125emX}}

\makeatletter
\newcommand{\linebreakand}{%
  \end{@IEEEauthorhalign}
  \hfill\mbox{}\par
  \mbox{}\hfill\begin{@IEEEauthorhalign}
}
\makeatother

\begin{document}

\title{The Quantum Education Ecosystem: A Review of Global Initiatives, Methods, and Challenges\\
}

\author{
\IEEEauthorblockN{Sara Ayman Metwalli\IEEEauthorrefmark{1},
Aryan Iliat\IEEEauthorrefmark{3},
Steven Thomas \IEEEauthorrefmark{5},
Suresh Nair \IEEEauthorrefmark{6},
Zizwe A. Chase\IEEEauthorrefmark{2},
Russell R. Ceballos\IEEEauthorrefmark{4}\IEEEauthorrefmark{7}
}

\IEEEauthorblockA{\IEEEauthorrefmark{1}Quantum Software Lab, The University of Edinburgh, Edinburgh, Scotland\\
smetwall@ed.ac.uk}

\IEEEauthorblockA{\IEEEauthorrefmark{2}Department of Electrical and Computer Engineering, University of Illinois Chicago, Chicago, USA\\
chase8@uic.edu}

\IEEEauthorblockA{\IEEEauthorrefmark{3}School of Physics and Applied Physics, Southern Illinois University, Carbondale, USA\\
aryan.iliat@siu.edu}

\IEEEauthorblockA{\IEEEauthorrefmark{4}Department of Physical Sciences, Olive-Harvey College -- City Colleges of Chicago, Chicago, USA\\
rceballos9@ccc.edu}

\IEEEauthorblockA{\IEEEauthorrefmark{5}Michigan State University, Michigan, USA\\
deshawn@msu.edu}

\IEEEauthorblockA{\IEEEauthorrefmark{6}Ina-Solutions, Tysons Corner, 
VA, USA\\
suresh.nair@ina-solutions.com}

\IEEEauthorblockA{\IEEEauthorrefmark{7}QuSTEAM Initiative, Iowa City, USA}
}

\maketitle

\begin{abstract}
Quantum information science and engineering (QISE) is advancing rapidly, creating an urgent demand for a quantum-literate, technically proficient workforce. Despite this need, quantum education initiatives remain fragmented across regions, educational levels, and instructional approaches, which constrains their scalability and overall impact. This paper offers a structured analysis of the current quantum education ecosystem by synthesizing global initiatives, pedagogical strategies, and emerging trends.

Quantum education is examined through a dual framework that considers both learner progression and instructional methodology, emphasizing the evolution of educational approaches from conceptual exposure to formal reasoning and practical application. Analysis of data from international programs and academic literature reveals key challenges, including inequitable access, absence of standardized curricula, limited empirical evaluation, and discontinuities between educational stages.

Quantum education is more accurately conceptualized as a non-linear ecosystem rather than a traditional pipeline, characterized by multiple entry points, feedback mechanisms, and critical transition gaps. Based on this perspective, directions are proposed for developing more coherent, inclusive, and scalable educational frameworks that align with workforce requirements and technological progress.

This work presents a unified perspective on the quantum education landscape and outlines actionable strategies to enhance global quantum literacy and workforce preparedness.
\end{abstract}


\section{Introduction}
\label{sec:intro}

Quantum information science and engineering (QISE) is rapidly transforming diverse fields, including cryptography~\cite{gisin2002quantum}, secure communication, materials science~\cite{bauer2020quantum}, and computation. As governmental and industrial investments increase, the demand for a workforce proficient in quantum technologies also rises. Nevertheless, the advancement of quantum education remains fragmented, unevenly distributed, and insufficiently aligned with workforce requirements.

While several universities and online platforms now offer introductory quantum courses, these opportunities are often limited to learners with an advanced background in mathematics or physics. Consequently, younger students, educators, and individuals from non-STEM disciplines encounter significant barriers to participation. The majority of existing academic programs remain concentrated at the graduate level, where instruction emphasizes quantum computing applications over theoretical principles or interdisciplinary integration. Although some institutions have recently established bachelor’s degrees in quantum information science, these initiatives are still geographically and institutionally constrained.

To bridge this gap, numerous global initiatives have emerged to broaden participation and enhance accessibility, spanning K–12 outreach, undergraduate curriculum development, and workforce training. Programs such as the Q-12 Education Partnership, QuSTEAM, DigiQ, QTIndu, and IBM’s Asia Quantum Education Network demonstrate a growing recognition that quantum education should be interdisciplinary, inclusive, and linked to real-world applications. Despite these efforts, many initiatives operate independently, leading to a lack of cohesion across educational stages and instructional methods.

This paper examines the current quantum education ecosystem through two primary perspectives: learner age groups and instructional methodologies. By synthesizing global initiatives and pedagogical strategies, we seek to provide a structured understanding of how quantum education advances from early conceptual exposure to workforce readiness. In this analysis, we identify key gaps, such as limited access, insufficient empirical evaluation, and discontinuities between educational stages, and highlight opportunities to develop a more coherent, scalable, and inclusive framework.

Specifically, this paper:
\begin{itemize}
    \item Surveys global quantum education initiatives across educational levels (Section~\ref{sec:one});
    \item Categorizes instructional approaches by learner stage and methodology (Section~\ref{sec:two});
    \item Identifies systemic gaps in accessibility, curriculum design, and evaluation (Section~\ref{sec:three});
    \item Proposes directions for developing a more integrated quantum education ecosystem (Section~\ref{sec:four}).
\end{itemize}

This perspective conceptualizes quantum education as a dynamic ecosystem characterized by multiple entry points, feedback mechanisms, and critical transition gaps, rather than as a linear pipeline.

\section{Global Overview of Quantum Education Initiatives}
\label{sec:one}

\begin{figure*}[t]
    \centering
    \includegraphics[width=0.8\textwidth]{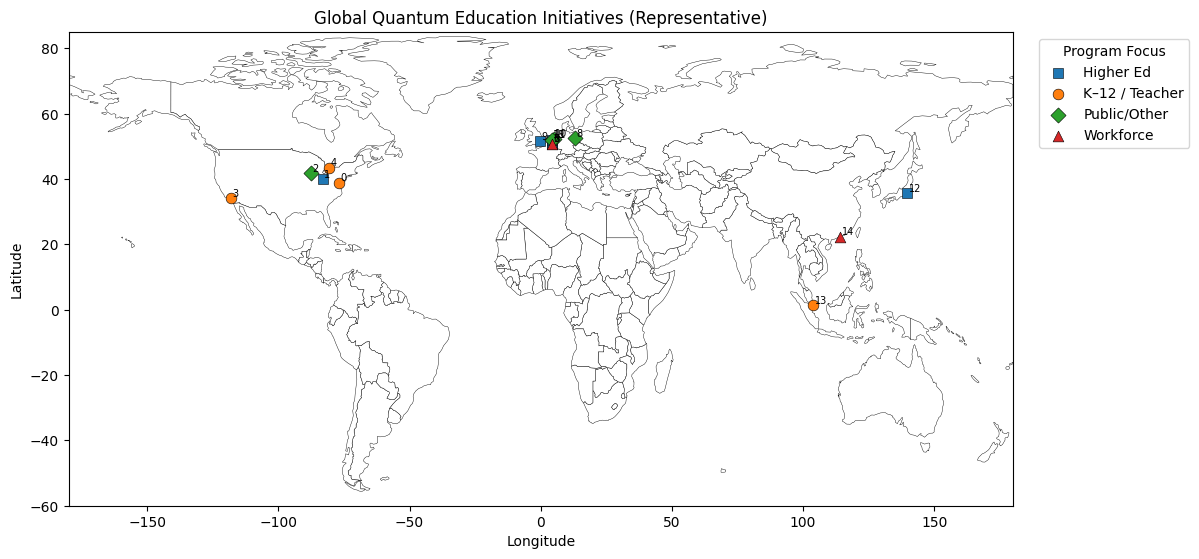}
    \caption{The distribution of some of the major quantum education/ workforce initiatives around the world.}
    \label{fig:map}
\end{figure*}

In the past decade, quantum education has transitioned from isolated academic efforts to a globally coordinated set of initiatives focused on developing a quantum-ready workforce. These initiatives operate at multiple levels, including national strategies, formal academic programs, industry partnerships, and open-access learning platforms. This evolution reflects the emergence of QISE as a distinct interdisciplinary field. Despite substantial growth in the number and diversity of initiatives, their structure remains fragmented, with limited coordination across regions, educational levels, and pedagogical approaches.

This section surveys representative efforts across regions, highlighting recurring structures and priorities that shape the global quantum education landscape. A distribution of some global initiatives is shown in Figure~\ref{fig:map}.

\subsection{Scope of Global Efforts}
Quantum education initiatives operate at various levels, from early exposure in K–12 classrooms to graduate-level specialization and professional upskilling. In the United States, programs such as the Q-12 Education Partnership and QuSTEAM develop curricular frameworks and expand undergraduate access. Organizations, including the Chicago Quantum Exchange (CQE) and Qubit by Qubit, engage learners across the entire educational spectrum. In Canada, the Institute for Quantum Computing (IQC) prioritizes teacher training through its Quantum for Educators program, supporting integration into secondary education.

In Europe, the Quantum Flagship coordinates large-scale educational efforts through initiatives such as DigiQ, which targets higher education, and QTIndu, which focuses on industry upskilling. Additional efforts include OpenHPI’s quantum MOOCs in Germany, offering open-access foundational training, and the United Kingdom’s National Quantum Technologies Program (NQTP), which integrates doctoral training with outreach activities. In the Netherlands, Quantum Delta NL advances public literacy through the "Quantum for Everyone" initiative and provides hands-on learning platforms via Quantum Inspire and QuTech Academy.

In the Asia-Pacific region, education networks such as IBM’s Asia Quantum Education Network and outreach programs led by Singapore’s Center for Quantum Technologies (CQT) foster regional collaboration and hands-on learning. Japan’s Quantum Technology Innovation Hubs (QIH) seek to standardize curricula and enhance industry–academic partnerships. Table~I provides a summary of representative initiatives across regions and educational levels.

In addition to formal and institutional programs, an expanding ecosystem of informal and community-driven initiatives has emerged. These include open-source platforms, online certification programs, hackathons, and global learning communities that offer alternative entry points into quantum education. Such initiatives are essential for broadening access and supporting non-linear learning pathways, especially for learners outside traditional academic environments.

Table \ref{tab:global_qed} summarizes some quantum education initiatives from around the world.

\begin{table*}[t]
\centering
\renewcommand{\arraystretch}{1.4}
\setlength{\tabcolsep}{10pt}
\resizebox{\textwidth}{!}{%
\begin{tabular}{|>{\centering\arraybackslash}m{3cm}|
                >{\centering\arraybackslash}m{4cm}|
                >{\centering\arraybackslash}m{4.5cm}|
                >{\centering\arraybackslash}m{4.5cm}|}
\hline
\rowcolor[HTML]{68CBD0}
\textbf{Region / Country} & \textbf{Program / Initiative} & \textbf{Level / Focus} & \textbf{Key Features / Highlights} \\ \hline

\rowcolor[HTML]{ECF4FF}
\multirow{4}{*} &
Q-12 Education Partnership~\cite{Q12Education2024} &
K–12, Teacher Training &
Develops quantum learning standards, teacher resources, and curricular materials. \\ \cline{2-4}

\rowcolor[HTML]{FFFFFF}
{\textbf{United States}} & QuSTEAM~\cite{brown2024accelerating} &
Undergraduate &
Modular interdisciplinary curriculum; scalable and inclusive design for diverse institutions. \\ \cline{2-4}

\rowcolor[HTML]{ECF4FF}
 & Chicago Quantum Exchange (CQE)~\cite{CQE2024}&
Multi-level (K–12 to Workforce) &
Regional hub linking universities, labs, and industry; supports internships, outreach, and events. \\ \cline{2-4}

\rowcolor[HTML]{FFFFFF}
 & Qubit by Qubit / The Coding School~\cite{QubitByQubit2024}&
K–12, Public &
Online global quantum courses in partnership with IBM; focuses on accessibility and diversity. \\ \hline

\rowcolor[HTML]{ECF4FF}
\multirow{1}{*}{\textbf{Canada}} &
IQC – Quantum for Educators~\cite{IQC2024} &
Teacher PD, High School &
Workshops and classroom toolkits to integrate quantum topics into secondary education. \\ \hline

\rowcolor[HTML]{FFFFFF}
\multirow{3}{*} &
Quantum Flagship Education \& Training &
Graduate, Workforce &
Pan-European competence framework with coordinated master’s and doctoral training. \\ \cline{2-4}
\rowcolor[HTML]{ECF4FF}
 {\textbf{Europe (EU-wide)}}  & DigiQ~\cite{DigiQ2024} &
Master’s / Higher Ed &
30+ open-access modules and 50+ master’s-level courses through the Digital Europe Programme. \\ \cline{2-4}

\rowcolor[HTML]{FFFFFF}
 & QTIndu~\cite{QTIndu2024}&
Workforce / Industry &
Industry-focused training to prepare companies and professionals for quantum adoption. \\ \hline

\rowcolor[HTML]{ECF4FF}
\multirow{1}{*}{\textbf{Germany}} &
OpenHPI Quantum MOOCs~\cite{OpenHPI2024} &
Public, Undergraduate &
Free open online quantum courses reaching over 17,000 learners globally. \\ \hline

\rowcolor[HTML]{FFFFFF}
\multirow{1}{*}{\textbf{United Kingdom}} &
National Quantum Technologies Programme (NQTP)~\cite{NQTP2024} &
Workforce, K–12 &
Includes Centres for Doctoral Training and “Quantum Ambassadors” outreach program. \\ \hline

\rowcolor[HTML]{ECF4FF}
\multirow{2}{*}{\textbf{Netherlands}} &
Quantum Delta NL – “Quantum for Everyone”~\cite{QuantumDelta2024}&
Public Literacy &
Free multilingual course with ITU and UNICC to promote global quantum literacy. \\ \cline{2-4}

\rowcolor[HTML]{FFFFFF}
 & Quantum Inspire / QuTech Academy~\cite{QuTechAcademy2024}&
Undergraduate, Public &
Online labs and tutorials for hands-on quantum programming using real backends. \\ \hline

\rowcolor[HTML]{ECF4FF}
\multirow{1}{*}{\textbf{Japan}} &
Quantum Technology Innovation Hubs (QIH)~\cite{QECJapan2024}&
University, Workforce &
National network standardizing curricula and fostering industry-academic collaboration. \\ \hline

\rowcolor[HTML]{FFFFFF}
\multirow{1}{*}{\textbf{Singapore}} &
Centre for Quantum Technologies (CQT) Outreach~\cite{CQTSingapore2024}&
K–12, Public &
Teacher workshops and public engagement events promoting quantum awareness. \\ \hline

\rowcolor[HTML]{ECF4FF}
\multirow{1}{*}{\textbf{Asia-Pacific (multi-country)}}&
IBM Asia Quantum Education Network~\cite{IBMAsia2024} &
University, Workforce &
Collaboration across Japan, Korea, India, and Singapore aims to train 40,000+ students. \\ \hline

\end{tabular}%
}
\vspace{0.5em}
\caption{Representative Quantum Education Initiatives Worldwide}
\label{tab:global_qed}
\end{table*}

\subsection{Patterns and Trends Observed}
Several consistent trends are evident across regions. First, there is a pronounced emphasis on early exposure and conceptual literacy. Programs such as Q-12, CQT Outreach, and Quantum for Everyone introduce quantum concepts using accessible narratives, visualizations, and classroom resources, thereby reducing barriers to entry for younger learners and the general public.

Second, there is growing alignment between education and workforce development. Initiatives such as QuSTEAM, QTIndu, and Quantum City prioritize transferable technical skills and interdisciplinary training, reflecting industry demand for expertise spanning physics, computer science, and engineering. These programs demonstrate a shift from purely theoretical instruction to application-oriented learning.

Third, online and hybrid learning modalities have substantially increased participation. Platforms such as OpenHPI, Quantum Inspire, and DigiQ provide global access to quantum education resources, reducing geographic and institutional barriers. These approaches are particularly important for scaling access and supporting lifelong learning, a defining feature of the quantum education landscape. Initiatives such as CQE in the United States, Quantum Delta NL in the Netherlands, and similar consortia demonstrate how localized networks of universities, government agencies, and industry partners can coordinate education, research, and workforce development. These ecosystems serve as scalable models but also highlight disparities in regions where such infrastructure is lacking.

\begin{figure}[t]
    \centering
    \includegraphics[width=\columnwidth]{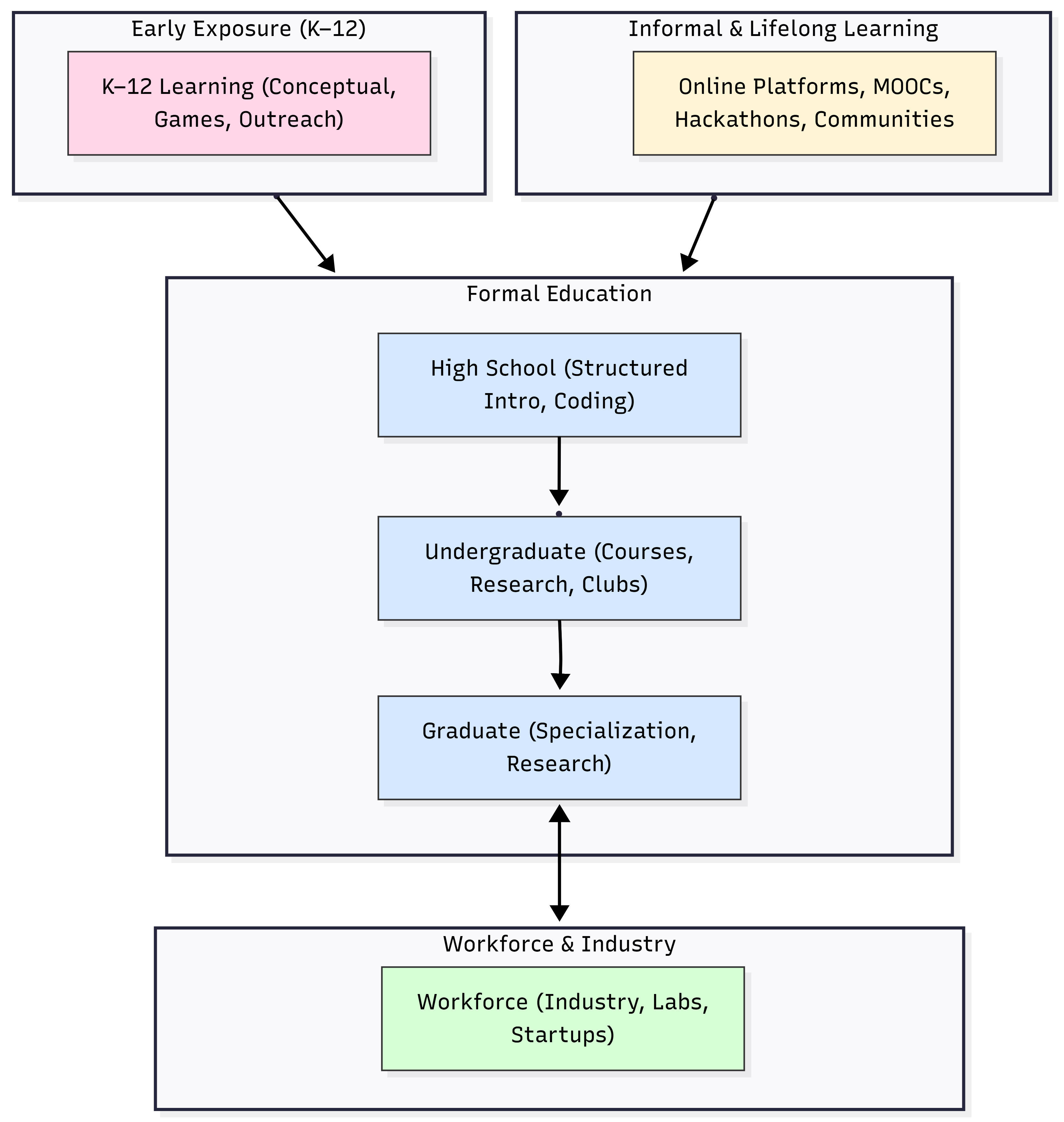}
    \caption{Education initiatives pipeline from early age to workforce readiness.}
    \label{fig:edu}
\end{figure}

\subsection{From Pipeline to Ecosystem}

Quantum education is frequently characterized as a linear pipeline from early exposure to workforce readiness. However, this perspective oversimplifies the complexity of learner pathways and institutional interactions. In reality, the quantum education landscape operates as a dynamic ecosystem marked by nonlinear progression, multiple entry points, and ongoing skill development.

Learners may enter through formal education, informal learning platforms, or professional training programs, and may transition between these pathways throughout their careers. Concurrently, feedback loops, especially between industry and higher education, play a critical role in shaping curricula, defining skill requirements, and guiding program development.

Figure~\ref{fig:edu} illustrates this ecosystem perspective, highlighting both the connections and discontinuities between different stages of quantum education. In particular, transitions between educational levels, such as the progression from high school to undergraduate study, represent critical bottlenecks where learners often disengage. These discontinuities underscore the need for more coherent, coordinated educational pathways.

Viewing quantum education as an ecosystem rather than a pipeline offers a more accurate framework for understanding current initiatives and for designing scalable, inclusive, and adaptable education strategies.

\section{Quantum Education at Different Levels}
\label{sec:two}

The complexity of quantum mechanics requires instructional approaches that vary across educational stages, leading to diverse strategies tailored to cognitive development, prior knowledge, and specific learning objectives. These stages constitute a developmental progression that guides learners from conceptual intuition to formal reasoning and application.

\begin{figure}[htbp]
\centerline{\includegraphics[scale=0.4]{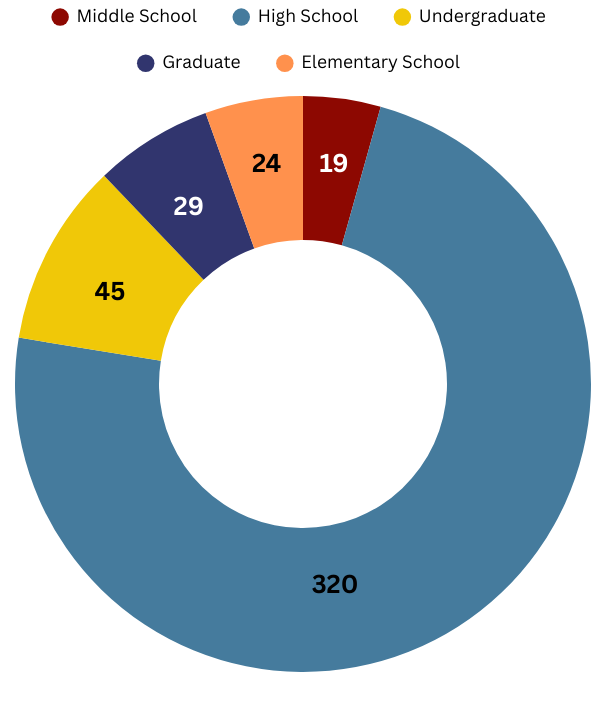}}
\caption{The distribution of quantum education papers on IEEE Explore targeting different age groups in formal education. 320 Papers targeting high school students, 19 for middle school, 24 for elementary school, 45 for undergrad, and 29 for graduate students in these schools.}
\label{cycle}
\end{figure}

A review of quantum education literature indicates that, among over 1,400 publications indexed in IEEE Xplore, only a subset explicitly addresses specific learner groups. Most of these works focus on high school students, while comparatively fewer target elementary, middle school, undergraduate, or graduate populations. Figure~\ref{cycle} illustrates this distribution, highlighting both the increasing interest in pre-university education and the uneven attention given to different educational levels.

\subsection{Elementary, Middle, and High School Education}
Recent efforts to introduce the basics of quantum computers to younger students focus on a higher level of abstraction to explain the concepts without using math. Instead, it can be achieved through games, interactive simulations, and simplified explanations of key concepts~\cite{evenbly2024exploring}. 

\subsubsection{Elementary School Quantum Education}  
Quantum education at the elementary level remains exploratory, with limited standardization and ongoing research into effective pedagogical methods. Instruction prioritizes conceptual exposure through storytelling, visual metaphors, and hands-on activities that introduce foundational concepts such as superposition and entanglement without relying on formal mathematics.
Programs such as Qubits for Kids~\cite{kuijpers2022qubits} and QPlayLearn~\cite{foti2021quantum} use tangible models and interactive narratives to foster early intuition. University and research institution outreach initiatives further support these efforts through guided activities and demonstrations. Although these approaches effectively promote curiosity and engagement, their long-term impact on conceptual understanding and progression remains unclear.

\subsubsection{Middle School Quantum Education}  

At the middle school level, quantum education adopts a more structured approach, incorporating basic principles and introductory mathematical reasoning. Programs developed by industry and educational organizations, including IBM Quantum, Qubit by Qubit, and Quantum for All~\cite{matsler2023quantum}, provide workshops, virtual labs, and guided activities tailored for early-stage learners.

Instructional methods at this stage frequently include interactive simulations~\cite{kohnle2015enhancing, migdal2022visualizing}, block-based programming environments~\cite{escanez2025qscratch, jacobs2013block}, and gamified learning experiences. Extracurricular activities such as quantum camps and escape rooms~\cite{QuantumEscapeRoomGermany, QuantumEscapeRoomIllinois} further promote engagement through problem-solving and collaboration. These approaches help bridge the gap between conceptual exposure and structured learning, although access remains inconsistent across regions and institutions.

\subsubsection{High School Quantum Education}  
High school represents a critical transition point in the quantum education pathway, where learners begin engaging with more formal representations of quantum concepts. Educational programs at this level introduce quantum circuits, algorithms, and programming through tools such as Qiskit and IBM Quantum Experience~\cite{angara2020quantum, angara2021teaching}.

In addition to coursework, students increasingly participate in hackathons~\cite{QWorldNYUAD}, summer programs, and research experiences that provide exposure to real-world quantum applications. Initiatives such as Chicago State University’s Quantum Sensing Summer Program~\cite{CSU_QuantumSensing} and hands-on platforms like the Quantenkoffer~\cite{qutools_Quantenkoffer} offer opportunities for experimental learning and early research engagement.

Despite these advances, the transition from high school to undergraduate quantum education remains a significant bottleneck. Many students lack sufficient mathematical preparation or structured guidance to progress, resulting in a critical discontinuity within the quantum education ecosystem.

\subsection{Undergraduate Education}

Undergraduate education is central to bridging foundational knowledge and specialized expertise in quantum information science and engineering (QISE). In recent years, universities have introduced more courses in quantum computing, quantum information, and quantum mechanics within physics, computer science, and engineering programs~\cite{economou2020teaching}.

Interdisciplinary degree programs and minors have emerged to address the need for both theoretical and applied knowledge~\cite{asfaw2022building}. These programs typically integrate linear algebra, quantum algorithms, and programming using platforms such as Qiskit and Cirq, preparing students for academic research and industry roles.

Experiential learning opportunities, including undergraduate research experiences (REUs)~\cite{linn2015undergraduate, buchanan2022current, russell2007benefits}, internships, and summer schools~\cite{malgierisummer}, have become essential components of undergraduate education. These experiences enable students to engage directly with quantum systems, software, and experimental setups.

Industry and nonprofit partnerships further enhance undergraduate learning. Initiatives such as the IBM Quantum Educators Program and QuSTEAM~\cite{QuSTEAM} provide curriculum resources, cloud access, and mentorship opportunities. Accelerated training programs and micro-credential certifications~\cite{barnes2025outcomes} also offer alternative pathways for learners seeking rapid entry into the field~\cite{perron2021quantum, perron2023educators, dzurak2022development}.

In fact, QuSTEAM was specifically highlighted in Nature Reviews:
\begin{itemize}
    \item “The QuSTEAM program helps faculty integrate QISE into their existing courses by providing detailed documentation for modular, customizable QISE curricula and by hosting professional development workshops for students, instructors, and administrators. Their curricula are designed to address many of the failings of traditional QISE pedagogy; introductory courses require no coding experience, and all courses emphasize project-based evaluations to broaden participation. The program also offers a wealth of professional development, networking, and mentorship opportunities with its network of QISE instructors and researchers — providing essential guidance for schools looking to build their own quantum programs,” Kayla Lee, IBM~\cite{lee2023faculty}.
\end{itemize}

Another modality important for expanding access to the QISE ecosystem is the "semi-formal" online educational platforms.  In addition to the ever-expanding set of science-based YouTube channels, there are online educational materials, sometimes as certification courses. One such example is the newly awarded NSF ExLENT award~\cite{NSF_BEGIN_2347101}, an industry-led, accelerated 6-week experiential learning program that serves as a micro-credential certification, enabling participants to gain hands-on skills through an online virtual classroom that introduces students from all backgrounds and interests to the fields of quantum computing and machine learning.  This provides an accelerated introduction to the field to provide students with the requisite skills to pursue more advanced certifications in the field (e.g. IBM Qiskit Certification).

Student-driven initiatives, such as quantum computing clubs and peer-led learning communities, contribute significantly to the ecosystem by fostering collaboration, self-directed learning, and community engagement. Alongside formal and institutional pathways, open-source and community-driven initiatives are playing an increasingly important role in undergraduate quantum education.

Organizations such as QWorld~\cite{qworld} and the Unitary Fund~\cite{unitaryfund} support global workshops, microgrants, and collaborative software development, enabling students to engage with quantum computing through hands-on, contribution-based learning. These pathways offer accessible entry points that complement traditional coursework and broaden participation beyond formal academic settings.

\subsection{Graduate and Professional Education}
At the graduate level, quantum education is highly specialized and research-driven. Programs at leading institutions such as Harvard, Caltech, and ETH Zurich focus on advanced topics, including quantum algorithms, quantum error correction, quantum cryptography, and experimental quantum systems.

Graduate students typically participate in hands-on research involving quantum hardware platforms such as superconducting qubits, trapped ions, and photonic systems. Funding agencies, including the National Science Foundation (NSF), the Department of Energy (DOE), and the European Quantum Flagship~\cite{riedel2017european}, support these efforts through fellowships and research grants.

Collaboration between academia and industry is particularly robust at the graduate level. Companies such as IBM, Google, Microsoft, and Rigetti provide access to hardware, internships, and joint research opportunities, enabling students to address real-world quantum challenges.

In parallel, quantum workforce development initiatives~\cite{raymer2019us, auffeves2022quantum, di2021cern, thew2019focus, purohit2024building} increasingly emphasize skills beyond technical expertise, such as entrepreneurship, policy, and technology translation. These programs reflect the evolving demands of the quantum industry and the necessity for interdisciplinary training.

The range of career pathways available to graduates in quantum technologies has expanded beyond traditional academic and research roles. Emerging positions include quantum software developers, applications scientists, hardware engineers, and quantum system technicians, as well as roles in policy, business development, and education. This diversification highlights the need for graduate training programs to align with both technical depth and transferable skills that support a broad and evolving quantum workforce.

One of the most important contributions in developing the quantum workforce and raising awareness about QISE is the Chicago Quantum Exchange (CQE)~\cite{ChicagoQuantum}. 

\subsection{Transitions and Continuity Across Levels}

Although each educational stage contributes uniquely to the development of quantum literacy and expertise, the connections between these stages remain inconsistent. The transition from high school to undergraduate education, in particular, represents a critical gap, as learners often encounter increased mathematical demands and diminished structural support.

Addressing this discontinuity requires intentional curriculum design, interdisciplinary bridging programs, and progressive learning experiences to support progression across educational levels. Integrating project-based learning, research exposure, and computational tools at earlier stages may help reduce this gap and establish more coherent pathways.

Collectively, these observations reinforce the need to conceptualize quantum education as an interconnected system, where progression, access, and continuity are actively designed rather than as a sequence of isolated stages.

\section{Quantum Education By Methodology}
\label{sec:three}

Although quantum computing concepts are consistent across different learner populations, instructional approaches differ substantially based on learners’ prior knowledge, cognitive development, and educational objectives. Rather than existing as isolated categories, these methods collectively constitute a developmental progression that guides learners from intuitive understanding to formal reasoning and practical application.

 As a cross-cutting approach that addresses each of the audiences referenced above, conceptual learning introduces the principles of QISE through analogies, visualizations, interactive narratives, and even games. Storytelling, thought experiments (e.g., Schr\"{o}dinger's), and classical comparisons help learners build an intuitive foundation before engaging with formal mathematics. Research indicates that analogical reasoning aids in initial comprehension, especially for young students and non-specialists~\cite{satanassi2021quantum}. However, misconceptions may arise when classical intuition is over-applied to quantum phenomena~\cite{majidy2024addressing}. 

This section categorizes widely used instructional approaches in quantum education and emphasizes their roles within a broader learning progression. As shown in Table~\ref{tab:sum}, each method presents distinct advantages and challenges and is most effective when incorporated into a learning pathway.

\subsection{From Intuition to Application: A Learning Progression}

Quantum education generally advances through multiple stages of engagement. Initial learning focuses on building intuition through conceptual explanations and visual representations. Subsequently, learners engage in interactive exploration by manipulating simplified models of quantum systems. As proficiency increases, learners construct and analyze quantum circuits using visual or block-based coding environments. Formal mathematical frameworks are introduced to provide rigor and generalization, culminating in hands-on programming and experimental interaction with quantum hardware.

This progression from intuition to application offers a valuable framework for understanding how various instructional methods contribute to learner development and how transitions between them can be effectively traced.

\subsection{Conceptual and Logical Learning}

Conceptual instruction introduces quantum ideas through analogies, storytelling, and non-mathematical reasoning. Programs such as Quantum for Everyone~\cite{QuantumForEveryone2025} and outreach modules from the CQT Outreach Program~\cite{CQTSingapore2024} exemplify this approach, using visual narratives and thought experiments to make abstract concepts accessible.

This method is especially effective for early learners and general audiences because it lowers barriers to entry and fosters curiosity. However, reliance on classical analogies may result in misconceptions when learners encounter formal quantum mechanics. Therefore, conceptual approaches are most effective when combined with subsequent instructional methods that refine and formalize understanding.

\subsection{Using Games to Teach Quantum}
Games provide students from a multitude of backgrounds and interests with a “safe space” to explore complex and technical topics they would otherwise avoid, allowing them to develop intuition.  This is essential in the field of QISE, where the principles and properties of the system are often counterintuitive.

The growth of game-based learning is becoming central to instruction and learning practices (Squire, 2010). Digital games offer players interactive contexts for thinking through and experimenting with complex problems hands-on. Using games as an educational method has been shown, in several contexts, to activate a wide range of thought styles, promote prosocial behavior, model alternate modes of action, and enable players to frame problems differently through procedural interaction~\cite{gee2008learning, gentile2009effects, belman2010exploring}.

The field of Games and Learning offers a powerful tool for engaging students in STEM-based learning, promoting asset development, and facilitating a transformative learning experience. Digital media, including social media, online games, and other shared technologies, when used to emphasize equity, full participation, social connection, and production, offer a powerful tool for building youth assets. The skills and learning patterns that students engage in when using digital media are self-motivated, creative, and social. Working in collaborative teams, addressing complex problems, learning through hands-on experience, and being mentored by caring, knowledgeable content experts can engender skills that help youth succeed in STEM fields.  Additionally, game design programs often reveal issues and topics relevant to designers, create artifacts that inform game design, and promote teamwork, collaboration, communication, and presentation skills.

To both "lower the bar for entry," as well as to establish an assessable platform in which students can develop intuition based on color and shape matching, two of the current authors have collaborated to develop and test prototypes of various quantum games (to be published at a future date) that introduce quantum coherence, entanglement, and interference in the simplest manner possible.  One such game reduces and simplifies quantum circuit construction and analysis to only using X and Z-basis representations for
\begin{itemize}
    \item $\vert0\rangle$, $\vert 1\rangle$, $\vert+\rangle$, and $\vert -\rangle$ states,
    \item Identity, X, Z, Hadamard, and CNOT operations,
    \item One and two-qubit measurement operations.
\end{itemize}
The game is both inspired by and leverages the basic principles and elements of diagrammatic calculus formulated within the framework of Quantum Picturalism~\cite{coecke2010quantum}.

Other examples of games that represent conceptual teaching efforts in the field include Quantum Chess~\cite{varga2024niel}, Quantum Tic-Tac-Toe~\cite{weingartner2023quantum}, and games used in high school quantum education~\cite{evenbly2024exploring}.  More in-depth reviews can be found herein~\cite{piispanen2023history, piispanen2025defining}.

\subsection{Interactive Simulations and Visual Learning}
Interactive platforms, such as IBM Quantum Composer and PhET simulations~\cite{phet}, allow students to see, interact with, and manipulate quantum states and observe the real-time effects of quantum phenomena without requiring programming skills. These tools provide an accessible way to introduce quantum superposition, entanglement, and measurement.
Studies suggest that visualization enhances understanding of counterintuitive quantum principles~\cite{zaman2021quantum, migdal2022visualizing}. However, the effectiveness of using no-code approaches (e.g., VQOL~\cite{VQOL}
, Quantum Flytrap~\cite{jankiewicz2022virtual} and VENUS~\cite{ruan2023venus}) is limited when learners do not transition to more formal, mathematical approaches~\cite{bangroo2023quaver} to solidify their learning and test their deeper understanding of the fundamentals of quantum computing.
These tools are most effective for high school and early undergraduate learners, bridging the gap between conceptual models and formal quantum mechanics. While they enhance engagement and reduce cognitive load, their scope is limited—often simplifying quantum systems to idealized two-level examples. Integrating these tools with guided reflection or teacher facilitation can significantly improve conceptual retention.

\subsection{Block-Based and Visual Coding Interfaces}
Tools like Qiskit Blocks~\cite{QiskitBlocks} and TinkerQubits~\cite{TinkerQubits} enable learners to build quantum circuits visually before transitioning to code-based quantum programming. These methods bridge the gap between conceptual learning and practical implementation.
Studies show that block-based coding (e.g., IBM Quantum Composer~\cite{loredo2025learn}) improves retention and lowers the entry barrier for quantum computing education~\cite{escanez2025qscratch}. Nevertheless, learners must eventually move to text-based coding (e.g., the Microsoft Quantum Development Kit for developers~\cite{hooyberghs2022introducing}) to better utilize hardware and deepen their understanding of how quantum computers work.
This approach is particularly suitable for late high school and early undergraduate learners who are developing computational literacy. While it provides a strong conceptual foundation for algorithmic thinking, a key challenge is ensuring a smooth progression to code-driven environments that reflect professional quantum programming workflows.

\subsection{Mathematical and Theoretical Approaches}
These approaches are often used with older learners, mainly after high school. While undergraduate and graduate courses typically introduce quantum computing through linear algebra, quantum algorithm design and implementation, and computational complexity theory, students are more often introduced to quantum mechanics via a position-based approach, rather than the spin-based approach typically favored by QISE educators~\cite{buzzell2025quantum}. Textbooks, such as \textit{Quantum Computation and Quantum Information}~\cite{nielsen2010quantum} and \textit{Classical and Quantum computation}~\cite{kitaev2002classical}, remain standard for theoretical learning.
This approach provides a solid foundation, but the cognitive load can be high for students without a strong mathematics background~\cite{hennig2024mathematical}.
University courses under initiatives such as \textit{QuSTEAM}, \textit{DigiQ}, and the UK’s \textit{NQTP Doctoral Training Centres} combine formal mathematics with application-focused modules. These programs develop the needed reasoning and analytical skills necessary for advanced research. However, the heavy emphasis on theory can deter learners without strong mathematical preparation. Modern curriculum models increasingly blend theory with visual or computational components, using simulation-based assignments to contextualize formal derivations.

\subsection{Hands-On Quantum Programming and Experimentation}
Hands-on learning through real or simulated quantum hardware has become central to undergraduate, graduate, and workforce education. Platforms like \textit{IBM Quantum}, \textit{Quantum Inspire}, and \textit{QuTech Academy} enable learners to execute circuits on actual devices, reinforcing connections between code and quantum state evolution. Laboratory-based programs—such as Canada’s \textit{Quantum City} and Japan’s \textit{Quantum Education Consortium}—extend this approach to hardware characterization and quantum sensing. These experiences cultivate practical skills in system calibration, data analysis, and error mitigation. The main challenges lie in limited hardware access and the rapid evolution of cloud platforms, which can make course materials quickly outdated.

\subsection{Experimental and Hardware-Centered Learning}
Beyond software interaction, advanced programs emphasize direct engagement with experimental platforms. Graduate students and professionals gain exposure to quantum optics, superconducting qubits, and cryogenic systems through internships or research fellowships. The European \textit{QTIndu} program and national centers under the \textit{Quantum Flagship} have developed modular training aligned with industrial needs. Such experiences provide the most authentic link between theory and application, though their cost and equipment requirements limit widespread implementation. Virtual laboratories and cloud-connected testbeds are emerging solutions that extend access to experimental learning worldwide.

\begin{table*}[t]
\centering
\renewcommand{\arraystretch}{1.4}
\setlength{\tabcolsep}{10pt}
\resizebox{\textwidth}{!}{%
\begin{tabular}{|>{\centering\arraybackslash}m{3.5cm}|
                >{\centering\arraybackslash}m{3.5cm}|
                >{\centering\arraybackslash}m{4.5cm}|
                >{\centering\arraybackslash}m{4.5cm}|}
\hline
\rowcolor[HTML]{68CBD0}
\textbf{Method} & \textbf{Best For} & \textbf{Advantages} & \textbf{Challenges} \\ \hline

\rowcolor[HTML]{ECF4FF}
Conceptual \& Logical Learning &
Beginners (K–12, general audiences) &
Builds intuition and early curiosity; accessible without prior math background &
Risk of misconceptions from oversimplified analogies \\ \hline

\rowcolor[HTML]{FFFFFF}
Interactive Simulations &
High school, early undergraduate &
Enhances visualization of superposition and interference; low entry barriers &
Limited depth; requires facilitation for deep understanding \\ \hline

\rowcolor[HTML]{ECF4FF}
Block-Based Coding &
Late high school, undergraduate &
Introduces quantum algorithms visually; bridges conceptual and programming stages &
Requires transition to text-based coding for advanced skills \\ \hline

\rowcolor[HTML]{FFFFFF}
Theoretical \& Mathematical Approach &
Undergraduate, graduate &
Develops needed reasoning and formal understanding; foundation for research &
Can be inaccessible to learners without strong math preparation \\ \hline

\rowcolor[HTML]{ECF4FF}
Hands-On Quantum Programming &
Undergraduate, graduate, workforce &
Offers authentic experience with real hardware; promotes applied skill development &
Limited hardware access; rapidly evolving software ecosystems \\ \hline

\rowcolor[HTML]{FFFFFF}
Experimental \& Hardware Learning &
Graduate, professional, research &
Provides deep understanding of physical systems; industry-relevant training &
High cost and limited equipment availability \\ \hline
\end{tabular}%
}
\vspace{0.5em}
\caption{Summary of Methods Used in Quantum Computing Education}
\label{tab:sum}
\end{table*}

\subsection{Integrating Methods Across the Learning Continuum}

The instructional methods described above are most effective when integrated into a coherent learning pathway. Conceptual and game-based approaches facilitate early engagement, while simulations and visual tools bridge understanding toward programming and formal reasoning. Mathematical instruction and hands-on experimentation subsequently enable deeper expertise and workforce readiness.

Designing effective quantum education programs requires not only selecting appropriate methods for specific audiences but also ensuring smooth transitions between them. Misalignment between instructional approaches, such as introducing formal mathematics without sufficient conceptual grounding, can contribute to learner disengagement and reinforce existing gaps in the education ecosystem.

This progression further underscores the importance of interdisciplinary integration, in which computational, physical, and engineering perspectives are introduced simultaneously rather than sequentially.

\section{Key Gaps and Future Directions}
\label{sec:four}
Although quantum education initiatives have expanded rapidly worldwide, significant structural and pedagogical gaps persist, limiting scalability, inclusivity, and long-term impact. These challenges arise from misalignment across educational stages, instructional approaches, and workforce requirements. Addressing these issues necessitates a shift from isolated programs to coordinated, ecosystem-level design.

\subsection{Limited Early-Stage Integration}

Although interest in introducing quantum concepts at the K–12 level is increasing, most initiatives remain exploratory and are not integrated into standard curricula. Early exposure is essential for developing intuition and sustained engagement; however, current efforts are typically limited to extracurricular activities or short-term outreach programs.

Integrating age-appropriate quantum concepts into existing STEM curricula, particularly in physics, mathematics, and computer science, could establish a more sustainable foundation for future learning. The development of structured, research-informed materials for early education remains a critical area for ongoing research~\cite{danaci2023manqala}.

\subsection{Inequitable Access and Global Disparities}

Access to quantum education remains uneven across regions and institutions. Most well-established programs are concentrated in North America and Europe, while many regions—including parts of Africa, South America, and Asia—have limited access to resources, infrastructure, and trained educators.

Within well-resourced regions, access frequently depends on institutional capacity, geographic proximity to research centers, or prior STEM preparation. Although initiatives such as Q-12, DigiQ, and IBM’s Asia Quantum Education Network have expanded participation, broader adoption of open-access materials, cloud-based platforms, and international collaborations is necessary to reduce barriers.

Open-access quantum learning platforms and hardware-sharing programs could help bridge this gap~\cite{meyer2024disparities}. Research labs and industries (e.g., IBM) develop free online programs to educate and train learners in quantum computing, such as the Quantum Summer School~\cite{IBMQuantumChallenges} and the Qiskit YouTube channel~\cite{QiskitYouTube}.

Addressing these disparities requires attention to language accessibility, cultural context, and local capacity-building to ensure that quantum education efforts are both globally distributed and locally relevant.

\subsection{Lack of Empirical Evaluation and Pedagogical Research}

Many quantum education initiatives report high levels of participation and engagement; however, systematic evaluation of learning outcomes is lacking. Few programs incorporate longitudinal studies or standardized assessment tools to measure conceptual understanding, skill development, or progression across educational stages.

The development of validated assessment instruments, including concept inventories, competency-based evaluations, and longitudinal tracking frameworks, would facilitate more rigorous comparisons across programs and support evidence-based curriculum design. Strengthening the research foundation of quantum education is essential for identifying effective practices and scaling successful models.

\subsection{Discontinuities Between Educational Stages}

A significant challenge within the current ecosystem is the lack of continuity between educational levels. The transition from high school to undergraduate education is a major bottleneck, as learners frequently encounter increased mathematical rigor without adequate conceptual or computational preparation.

This middle gap reflects a broader issue: educational stages are often designed independently, lacking coordination of learning objectives, instructional methods, or expected competencies. Bridging this gap requires concrete curricula, interdisciplinary project-based learning, and early exposure to computational and mathematical tools to support progression. For example, ManQala provides a rich platform for designing control protocols relevant to quantum technologies~\cite{danaci2023manqala}, entanglement distribution over quantum networks~\cite{Chase2024SeQUeNCe} and so forth.

\subsection{Curriculum Coherence and Skill Alignment}
Another ongoing challenge involves the lack of a consistent framework defining the knowledge and competencies required for quantum literacy and workforce readiness. Curricula vary widely in depth and focus—some emphasize conceptual intuition, others coding or laboratory skills—leading to fragmentation across educational stages.  
While frameworks such as QuSTEAM and QTIndu have begun articulating transferable competencies, broader coordination is needed between academia, industry, and government agencies to map learning outcomes to workforce requirements.  
Developing an internationally recognized “quantum competency framework” could help standardize course objectives and clarify progression pathways from high school through professional certification. Hence, each country\/ area would need a set of requirements. However, the fundamental concepts that should be covered in any standardized curriculum would probably be the same~\cite{lee2023education}. 

\subsection{Teacher Preparation and Professional Development}

The effectiveness of quantum education depends heavily on educators; however, most teachers lack formal training in quantum information science. Existing professional development programs, such as IQC’s Quantum for Educators and Q-12 workshops, offer valuable resources but are limited in scale.

Expanding teacher training through scalable models—such as online certification programs, regional workshops, and peer-learning communities—would improve confidence and capacity among educators. Supporting teachers as active participants in the quantum education ecosystem is essential for long-term sustainability.

\subsection{Infrastructure, Collaboration, and Resource Sharing}

Access to quantum hardware, software platforms, and laboratory environments remains uneven. While cloud-based platforms such as IBM Quantum and Quantum Inspire have expanded access, hands-on experimental training is still largely confined to well-funded institutions.

Developing shared infrastructure, interoperable platforms, and remote laboratory environments could help democratize access to advanced quantum tools. Collaborative models that pool resources across institutions and regions may also reduce duplication and improve efficiency.

\subsection{Future Directions}

Addressing these challenges requires a shift from isolated initiatives to coordinated ecosystem-level strategies. As illustrated in Figure~2, quantum education is inherently non-linear, involving multiple entry points, feedback loops, and transitions across formal and informal learning environments.

Future efforts should prioritize:
\begin{itemize}
    \item \textbf{Equity and inclusion:} Expanding access through open resources, global partnerships, and targeted outreach;
    \item \textbf{Evidence-based design:} Integrating assessment and educational research into program development;
    \item \textbf{Curriculum coherence:} Aligning learning outcomes across educational stages and institutions;
    \item \textbf{Workforce alignment:} Defining and mapping competencies to emerging quantum career pathways;
    \item \textbf{Educator support:} Scaling teacher training and professional learning communities;
    \item \textbf{Infrastructure sharing:} Leveraging cloud and collaborative platforms to expand access to tools and experimentation.
\end{itemize}

Together, these directions emphasize the need to design quantum education as a connected, adaptive system rather than a collection of independent efforts. Such an approach is essential for building a sustainable, inclusive, and globally distributed quantum workforce.

\section{Conclusion}
\label{sec:conclusion}

Quantum education is rapidly becoming a foundational element within the broader quantum technology ecosystem. The initiatives reviewed in this paper demonstrate significant progress in raising awareness, developing curricula, and linking learners to workforce pathways across various educational levels. From early conceptual exposure in K–12 settings to specialized graduate training and professional development, the global landscape exhibits growing momentum and diversity in approaches to teaching quantum information science and engineering (QISE).

Despite this progress, persistent structural challenges remain. Educational initiatives are fragmented across regions, institutions, and learner stages, with limited coordination among pedagogy, curriculum design, and workforce requirements. Transitions between educational levels, particularly from high school to undergraduate study, represent significant bottlenecks. Additionally, inequitable access and the absence of empirical evaluation continue to constrain scalability and inclusivity.

Analysis of quantum education through the combined perspectives of learner progression and instructional methodology underscores the need to move beyond a linear pipeline model toward a more dynamic, interconnected ecosystem. Effective quantum education requires both diverse instructional approaches and deliberate alignment across educational stages, disciplines, and learning environments.

The development of a sustainable quantum workforce will require coordinated efforts that integrate accessible curricula, educator training, empirical assessment, and global collaboration. Designing quantum education as an adaptive, inclusive, and evidence-based ecosystem is essential to ensure that learners can meaningfully engage with and contribute to the rapidly evolving quantum landscape.

\section*{Acknowledgment}
This paper is supported by the ASCR Award DE-SC0024096. Partial support for this project is provided for by the U.S. Department of Energy, Office of Science, ASCR ReACT-QISE Consortium (DE-SC0024096)
\bibliographystyle{unsrt}
\bibliography{bib}

\end{document}